\documentclass[a4paper]{article}
\usepackage{graphicx}
\usepackage{chicago}
\newcommand{\be}{\begin{equation}}
\newcommand{\ee}{\end{equation}}
\newcommand{\bd}{\begin{displaymath}}
\newcommand{\ed}{\end{displaymath}}
\newcommand{\bea}{\begin{eqnarray}}
\newcommand{\eea}{\end{eqnarray}}
\newcommand{\bi}{\begin{description}}
\newcommand{\ei}{\end{description}}
\newcommand{\bq}{\begin{quote}}
\newcommand{\eq}{\end{quote}}

\def\fo{\footnote}

\def\r{\rho}
\def\a{\alpha}
\def\b{\beta}

\def\g{\gamma}

\def\e{\epsilon}

\def\th{\vartheta}

\def\m{\mu}
\def\n{\nu}

\def\l{\lambda}
\def\L{\Lambda}
\def\s{\sigma}

\def\p{\varphi}

\def\Z{{\sf Z\kern-4.5pt Z}} 
\def\R{{\sf R\kern-5.0pt I}}
\def\Q{{\sf C\kern-5.0pt Q}}

\begin{document}
\bibliographystyle{chicago}
\twocolumn[
\author{Alexander~Unzicker\\
        Institute for Meteorology\\  
        University of  Munich\\[0.6ex]
        Theresienstr. 37,  D-80799 M\"unchen, Germany\\
{\small{\bf e-mail:}  aunzicker@lrz.uni-muenchen.de }}
\title{What can Physics learn from Continuum Mechanics~?}
\date{November 20th, 2000}
\maketitle
\begin{center}
Dedicated to Prof. Ekkehart Kr\"oner
\end{center}

\begin{abstract}
This paper is mostly a collection of ideas already published by
various authors, some of them even a long time ago. 
Its intention is to bring the reader to know some
rather unknown papers of different fields that merit interest 
and to show some relations between them the author
claims to have observed. In the first section, some comments 
on old unresolved problems in theoretical physics are collected. 
In the following, I shall explain what relation exists between
Feynman graphs and the teleparallel
theory of Einstein and Cartan in the late 1920s, and the relation of both
to the theories of the incompressible aether around 1840. 
Reviewing these developments, we will  have a look at 
the continuum theory of dislocations developed by Kr\"oner
in the 1950s and some techniques of differential geometry and topology
relevant for a modern description of defects in continous media.
I will then illustrate some basic concepts of 
nonlinear continuum mechanics and discuss applications
to the above theories. By doing so, I hope to attract attention to
the possible relevance of these facts for `fundamental'  physics.
\end{abstract}
\vspace{1.0cm}]

\tableofcontents

\section{Old unresolved problems in 
theoretical physics} 

After the foundation of modern physics with its cornerstones
quantum mechanics and general relativity up to 1930,
theoretical physics has
developed in a less revolutionary manner in the past 
decades.
Richard Feynman mentioned in his Nobel
lecture (1965)
 that he was driven by the hope to {\em calculate\/}
the rest energy of an electron - which is an experimentally
well-known quantity of 0.511 MeV. Not much has happened
yet towards the solution of this problem. 
Another example where physicists seem to
have surrendered regards the mass ratios $m_p/m_e = 1836.15...$,
$m_n/m_e= 1838.68...  $, $m_{\m}/m_e = 206.768...$ of protons, neutrons, 
myons and others with respect to the electron. 
 It seems like an evidence of incapacity for
present day physics that the only attempts to calculate these
numbers are some playing with powers of $e's$ and 
$\pi's$ without any physical background.\footnote{How to play at 
least efficiently, one can read in \citeN{Bai:89}.}
 The same development has taken place 
with the fine structure constant
$\a = 137.03597...$. Feynman commented:
\bq
 `It is one of the 
{\em greatest\/} damn mysteries of physics. 
We know what kind of a dance to do experimentally to measure
this number very accurately, but we don't know what kind of a
dance to do on a computer to make this number come out -
without putting it in secretely~! ... 
all  good theoretical
physicists put this number on the wall and worry 
about it'.\footnote{\citeNP{FeyQED}, chap.~4.}
\eq
Today's physicists seem to prefer announcing some great unification
now and then 
instead. Even more remote to a solution 
is the problem\fo{This problem, which was already 
mentioned by \citeN{Dir:39},
was recently put in evidence by \citeN{Wei:99}.}  of the
ratio of the electromagnetic and gravitational force, which
is around $10^{40}$.
In this sense, physics hasn't moved closer towards a great unification
since the speculations of  \citeN{Edd} and  \citeN{Dir:39}. 
But should physicists disregard these problems forever~?

Is there
a reason why nature does not permit us to
resolve these puzzles as a matter of principle,
 like the quadrature of the circle~?
If this should be the case,  
physicists havn't done their homework yet by proving 
 their `$\pi$' to be transcendent.

Quantum electrodynamics had a great success after having calculated the 
the magnetic moment $1.00115965 \m_B$ of the electron
and the Lamb shift of about 1040 MHz.
However, in view of the 
above unsolved questions\footnote{As Feynman pointed out in his
famous lectures \cite{FeyII}, the quantization of
electrodynamics did not resolve the basic 
inconsistency of electrodynamics that predicts (with
Coulomb's law 
 and the energy density of the field) an infinite 
energy for the electron.}, does this justify
to build a general theory of physics on renormalization~?
Even Feynman himself was never convinced of the 
correctness of that theory:
\bq
`It's surprising that the theory still hasn't been proved 
self-consistent one way or the other by now; I suspect that
renormalization is not mathematically legitimate.'
(\citeNP{FeyQED}, p.~128).
\eq

Dirac expressed himself more drastically:
`This is just not sensible mathematics. 
Sensible mathematics involves
neglecting a quantity when it turns out to be 
small - not neglecting it
because it is infinitely great and you do not want it.'
\footnote{cited by \citeN{Kak}, p.~12.}
However, since the times of QED a kind of monoculture 
of physical theories have been developed on its conceptual
basis.
Even Feynman commented self-ironically on the theories
based on QED: 
\bq
`so when some fool physicist gives a lecture 
at UCLA in 1983 and says: "This is the way it works, and
look how wonderfully similar the theories are," it's not because 
nature is {\em really\/} similar; it's because the physicists have
only been able to think of the same damn thing, over and over 
again.' (\citeNP{FeyQED}, chap.~4, p.149).
\eq

Is it therefore not astonishing that general 
relativity remained `off side' from the `rest' of theoretical
physics.\fo{Interestingly, experimenters face an embarrassing uncertainty
($1,5 \ 10^{-3}$) arising from discrepant measurements of the value of 
the gravitational constant $G$. Recently, \shortciteN{Var:99} suspected
a theoretical reason for this.}
 
In Ryder's \citeyear{Ryd:85} book on quantum field theory we can read: 
`the quantisation of the theory is beset by great problems.'...
`in electrodynamics the field is an actor on the spacetime stage, 
whereas in gravity the actor becomes the spacetime stage itself.'
Then the comment follows:
..`In view of this, the particle physicist is justified in 
ignoring gravity -  and because of the difficulties
mentioned above 
 is happy to!' This kind of reasoning
seems to be searching a key in the shine of a lamppost because 
one can see there, even if the key had been lost elsewhere in the dark.

In view of the unresolved problems physicists are rather justified to
dig out and reconsider old ideas of excellent scientists - 
and because of the pleasure of reading their papers they
should be happy to!

\section{Lorentz symmetries in elastic solids}

\subsection{Frank's discovery}

This section
should guide the readers attention to a paper 
`On the Equations of Motion of Crystal Dislocations' of
\citeN{Fra:49}. The abstract follows:
\bq
`It is shown that when a Burgers screw dislocation moves with velocity
$v$ it suffers a longitudinal contraction by the factor 
$\sqrt{1-\frac{v^2}{c^2}}$, where $c$ is the velocity of transverse
sound. The total energy of the moving dislocation is given by th
formula $E=E_0 /(1-\frac{v^2}{c^2})^{\frac{1}{2}}$, where $E_0$
is the potential energy of the dislocation at rest. Taylor
dislocations behave in a qualitatively similar manner,
complicated by the fact that both longitudinal and transverse
displacements and sound velocities are involved.'
\eq

A visualization of dislocations in crystals is given 
in fig.~\ref{disloc} in section~\ref{tool}.
As has been pointed out by Frank, dislocations appear to have 
a particle- like behaviour. Their motion in
a crystal close to the velocity of transverse sound is analogous
to the motion of a particle close to the speed of light.
Is this just a coincidence~?
At the first look, dislocations are a very special kind of defect. 
For deriving the above result, Frank considers the deformation field 
of a screw dislocation
\begin{equation}
u_x=0, \quad u_y=0 \quad u_z=(b/2 \pi)\arctan{\frac{y}{x}},
\label{sd}
\ee
which is, since \mbox{$\frac{\partial^2}{\partial x^2} u_z
+\frac{\partial^2}{\partial y^2} u_z=0$} and 
$div \ {\bf u}= \frac{d}{dz} u_z=0$, a statical solution of pure shear type
of  the Navier equation
\be
 - (\l + 2 \m) 
\ grad \ div \ {\bf u} +\m \ curl \ curl \ {\bf u}
= \r \frac{\partial^2 {\bf u}}{\partial t^2}
\label{navier}
\ee
the displacement components ${\bf u} =(u_x, u_y, u_z)$ have to satisfy.
For the following, it is sufficient to consider
only the last two terms.

A dislocation propagating in $x$-direction with velocity $v$
must be represented by a time-independent 
function of $x'$, $y$ and $z$, where
$x' = x-v  t$. With this substitution, $\frac{\partial^2}{\partial t^2}$
becomes $v^2 \frac{\partial^2}{\partial {x'}^2}$, and the
remaining terms of eqn.~\ref{navier} read:
\be
\m (\frac{\partial^2}{\partial y^2}+\frac{\partial^2}{\partial z^2})+
(\m -v^2 \r ) \frac{\partial^2}{\partial {x'}^2}=0
\label{navier2}
\ee

With the further substitution 
\be
x'' = x' \sqrt{1-\frac{v^2 \r}{\m}}
= (x- v  t) (1-\frac{v^2}{c^2})^{-\frac{1}{2}},
\ee
where $c=\sqrt{\m/ \r}$
is the velocity of transverse sound, the solution of a propagating
dislocation has a form identical to the solution 
(\ref{sd}) of the dislocation  at rest, apart from the
substitution $ x \rightarrow x'$, a `Lorentz contraction'.

As \citeN{Wee:79}, p.~8, commented:
`This distorsion is analogous to the contraction and 
expansion of the electric field surrounding 
an electron.'
Frank went ahead and showed that the elastic energy
of the moving screw dislocation increases with the 
factor \mbox{$(1-\frac{v^2}{c^2})^{-\frac{1}{2}}$}.

The question
arises, whether this relativistic behaviour\footnote{{\bf 
Radiation damping.}
Interestingly, a phenomenon of radiation damping seems to
occur in the dynamics of dislocations (\citeNP{Kos:62}; 
\citeNP{Kos:79}). That means, a part
of the energy used in order to accelerate a dislocation
is dissipated by the production of transversal sound waves.
It should be noted that radiation damping in classical
physics is everything but well understood. As e.g. \cite{Dir:38b}, eq.~24,
\shortciteN{LanII}, par.~75 or \shortciteN{FeyII}, 
chap.~28 point out, the Lorentz force
$\vec F_L = e \vec v$ x $\vec B$ is just an 
approximation for small values of $dv/dt$, and a
general formula for the radiation emitted by an
accelerated electron does not exist. The quantization
of electrodynamics didn't resolve this problem either.}
is a consequence of
the special solution (\ref{sd}) or a more general effect. 
However, things would get more complicated only if the
dilatational part $div \ {\bf u}$ of the displacement 
in eqn.~\ref{navier} does {\em not\/} vanish. If it vanishes instead,
the above transformation $x \rightarrow x''$ can obviously be applied to
{\em every\/} solution of (\ref{navier}). The condition
of the vanishing dilatation can be formally realized by letting
go $\l$ to infinity, which does physically mean that the medium is 
incompressible.\footnote{Of course, we are dealing with linear elasticity
and have tacitly assumed small displacements. The nonlinear issues
are discussed below.} The reader who is interested in details may
look how other authors like \citeN{Esh:49}, \citeN{Wee:79},
\citeN{Gun:88}, and
in a somewhat redundant
way, \citeN{Gun:96} have developed
these analogies further, obtaining all features of special relativity
including time dilatation etc. Thus,
propagating solutions in an incompressible elastic continuum
behave exactly as relativistic particles, if the speed 
of light is identified with the velocity of transverse sound.

\subsection{MacCullagh's theory}
Can these relativistic effects, 
apart from being a curiosity of 
elasticity theory, have a deeper meaning~? It does not seem so,
because all attempts of describing fundamental physics with
continuum mechanics in the 19th century have been
falsified by the famous experiments by Michelson and Morley 
that seem to have disproved the concept of an aether. 
What's wrong here~? The point is, the  physicists of the 19th
century imagined particles as made of an external substance 
distinct from the `aether', which for some reason 
can pass through it without (or with infinitely
little) friction even with  infinitely
great velocities. 

The aether theorists Young and Fresnel, Stokes, Navier, 
Cauchy, Lord Kelvin and Green 
never thought of particles as being topological defects
 creating a displacement field - which is not astonishing,
since the first examples of such defects, dislocations 
in solids, were discovered in \citeyearNP{Tay:34}  
by Taylor, almost 30 years after aether theories
had disappeared from the stage of theoretical physics.
In view of the results of \citeN{Fra:49} and others, however, one
must say that {\em the wrong concept was not describing spacetime
as an elastic continuum but a wrong or missing picture of 
particles moving in it\/}.
 Therefore, theories of an incompressible
aether like for example the one of \citeN{MacC:39} do not
contradict the the experiments of Michelson and Morley,
if moving particles are assumed to be propagating 
solutions.\fo{As has been pointed out by \citeN{Dir:51},
aether theories do not contradict quantum mechanics either,
as long as the absolute velocity of the aether material
appears as a nonmeasurable quantity.} 

Let's have a look how MacCullagh\footnote{This was the result presented 
in 1839 to the Royal Irish Academy by MacCullagh and published in
1848 in Trans.Roy. Irish Acad. xxi, p.17. The interested
 reader is referred to the excellent
review on aether theories by \citeN{Whi:51}, p.~142. ff; p.~280}
 identified the elecromagnetic
quantities with those of an incompressible elastic aether:

According to this theory, one 
may identify the electric\fo{To be precise, the 
electrostatic induction $\vec D$, which 
has to be divided by the dielectricity constant $\e_0$ to obtain 
$\vec E$.} field $\vec E$ with the curl of the
displacement field $\nabla$ x ${\bf u}$ and the magnetic 
field strength $\vec H$ with its 
time derivative $\frac{d {\bf u}}{d t}$.
Then, the Navier equation (\ref{navier}) reduces to
\be
\r \frac{\partial^2 {\bf u}}{\partial t^2}=\mu \ curl \ curl \ {\bf u}
\label{inc}
\ee
which is ($\m$ is the shear modulus) equivalent to Maxwell's 
\be
\m_0 \frac{\partial \vec H}{\partial t}=  \frac{1}{\e_0} \ curl \ \vec E,
\ee 
whereas $div \ \vec H=0$ follows directly from the incompressibility
condition $div \ {\bf u}=0$ which implies 
$div \  \frac{d {\bf u}}{d t}=0$. By definition
\be
div  \ curl  \ {\bf u} =0 \quad \mbox{and} \
curl \frac{d}{d t} {\bf u} =\frac{d}{d t} curl \ {\bf u}
\ee
holds, which correspond to Maxwell's second pair 
of equations {\em in vacuo\/}. \citeN{Whi:51}, p.~143, comments:
\bq
It is evident from this equation (eqn.~\ref{inc}) that if $div \ {\bf u}$
is initially zero it will be always zero; we shall suppose this to be 
the case, so that no longitudianal waves exist at any time in the 
medium. One of the greatest difficulties which beset elastic-solid
theories is thus completely removed.\fo{MacCullagh assumed furthermore
the elastic energy to be a function of $curl \ {\bf u}$. As we shall see 
later, this additional assumption is not necessary.}
\eq

Before I discuss in section~\ref{Cha} the topological issues 
how charged particles may enter in this model, I would like to point
out the connection with a paper of
Einstein in \citeyearNP{Einst:30}, where he states: `(46), (47) ...
correspond to Maxwell's equations of empty space.' Einstein does
not mention MacCullagh, although according to historians 
\cite{Kos:92} he had given up his rejection to ether
already around 1920. But it is better to tell the 
story from the beginning:

\section{Einsteins teleparallel theory}

\subsection{The early papers 1928}
There is an amazing contrast between the public 
admiration for Einstein for having developed general relativity
 and the importance
that is given to his later
work in differential geometry. \fo{To avoid confusion,
it should be mentioned that this theory distinguishes substantially
from the so-called Einstein-Cartan-Sciama-Kibble (ECSK) theory.
See \citeN{Hehl:76} for a review of several theories including
torsion.}
More than once I happened to hear the statement
that after 1920  Einstein had 
published just nonsense. Symptomatically,
his work on teleparallel geometries 
has not been translated  in English yet.\fo{A translation of
some of his papers is availible under 
\mbox{{\em www.lrz.de/\~ \ aunzicker/ae1930.html\/}}.}
Of course, there is a reason for the 
disregard of Einstein's work of the
years after 1920: the conflict he had with 
quantum mechanics. His continuous objections,
for example at the Solvay conference in 1927,
could not unsettle the success of the new 
theory. On the contrary: people were
realizing more and more that quantum mechanics was a 
good physical theory because it  described
 the experiments, and got tired of the
philosophical attacks launched by Einstein.
In plain words: Einstein was a nuisance in the 
1920s, and it is quite understandable that physicists 
were annoyed of the work that he proposed as
an alternative to quantum mechanics and called
unified field theory. Thus, why should we deal
with his cumbersome tensor calculus developed in
a series of papers\footnote{\citeN{Einst:28};
 \citeN{Einst:28a}; \citeN{Einst:30}.}
and follow all his attempts that at the end were 
discarded by Einstein himself~?
A closer look at these 
 papers, however, reveals that there is
not only no contradiction to quantum mechanics on a 
conceptual level\footnote{Of course,
the formalism is quite different from that
of quantum mechanics, as that of GR is.},
but  there even arise some surprising
facts from that geometry that remind us from 
the quantum behaviour of particles. I will discuss that
in section~\ref{QB}.
But let's listen to \citeN{Einst:28} now:
\bq
`Riemannian Geometry has led to a physical description of the 
gravitational field in the theory of general relativity, but
it did not provide concepts that can be assigned to the 
electromagnetic field. Therefore, theoreticians aim to 
find natural generalizations or extensions of Riemannian
geometry that are richer of concepts, hoping to get to
a logical construction that unifies all physical field
concepts under one single leading point.'
\eq

\subsection{Torsion in Riemannian geometry}
Einstein was convinced that the geometric description
of physics does not stop at the rather special case
of Riemannian geometry.\footnote{It should be mentioned that
the notion of  Riemannian geometry seems to have changed. Einstein
intended a geometry in which the connection was determined by
the metric only, with the absence of torsion (see also 
\citeNP{Sch:54},\shortciteNP{Bil:55}). Modern texts like
\citeN{Nak:95} instead require just the existence of a Riemannian
metric.}
 In a later paper \cite{Einst:30}, he says:
\bq
`To take into account the facts
(...) gravitation, we assume the existence of 
Riemannian metrics. But in nature we also have electromagnetic fields, 
which cannot be described by Riemannian metrics. 
The question arises:
How can we add to our Riemannian spaces in a logically natural 
way an additional 
structure  that provides all this with
a uniform character~?'
\eq

In the following Einstein refers to an idea that Cartan 
had pointed out to him already in \citeyearNP{Cart:22a} - 
and Einstein did not understand at that
time-, the `Columbus connection'.\footnote{Connection is
the differential geometric entity that governs the
law of parallel transport of vectors.}
 For Columbus, navigating 
straight meant going westwards. In terms of differential 
geometry: parallel transport of vectors means keeping a
fixed angle to the lines of constant latitude, whereas 
usually the straight lines on a sphere are defined as
the great circles\fo{Therefore, it is necessary
to distinguish between {\em autoparallels\/}, on which
vectors remain parallel, and {\em extremals\/}, that 
maximize the covered distance 
.} (fig.~\ref{columb}).

\begin{figure}[h]
\includegraphics[width=60mm]{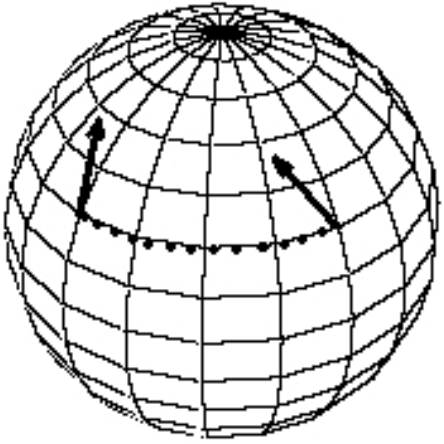}
\caption{Visualization of the connection (vector transport rule)
proposed by Cartan to Einstein.
While transporting a vector (along the dotted line) the 
angle with the meridians is kept fixed. Thus, directions may
be compared globally (Whenever we are speaking of `west', `east', 
`north' and `south', we are comparing directions globally~!).
If this is possible, a teleparallel
connection can be given to the manifold and
 the curvature tensor vanishes.}
\label{columb}
\end{figure}

Surprisingly, with this new connection the sphere has zero curvature
 but nonzero 
torsion.\fo{One can imagine best
 the difference between curvature and torsion
with differential forms. Both are 2-forms, that means quantities
that have to be integrated over a 2-surface. If one transports a vector
along a closed curve that bounds this surface, in the case of 
curvature it comes back {\em rotated\/}, and in the case of torsion
{\em shifted\/}. Because this shift is done by a vector, torsion
is called  a $vector$-valued form, whereas curvature could be
called a `rotation-valued' form. Correctly speaking, it is a
$Lie-algebra$- valued form. If the vector becomes 
just rotated (in the so-called metric-compatible case) the curvature 
form takes values in so(3), the Lie algebra of orthogonal 
rotations in three-dimensional space. For an introduction to differential
forms, see \citeN{Fla} or \citeN{Nak:95}.}
If one looks at Fig.\ref{columb}, it becomes clear what 
Einstein said:

\bq
`In every point there is a ... orthogonal $n$-bein\fo{$n$-`leg',
from German `bein', means $n$ orthogonal unit vectors.} . (...)
The orientation of this $n$-beins is not important in a Riemannian
manifold. We assume, that these (...) spaces are governed 
by still another direction law. We assume, (...) it makes sense
to speak of a  parallel orientation of all $n$-beins together (...).'
\eq

That was the idea that Einstein applied to space\-time-  describe the same 
physics with another differential geometric entity. Instead
of nonzero curvature and vanishing torsion he proposed
vanishing curvature and nonzero torsion - from the example
fig.~\ref{columb}
it should be clear that this does not change the geometry
of space. The advantage 
is that torsion in four dimensions has more components that
curvature - that means one can pay the bill for describing
gravity and hope that electromagnetism comes out of the additional components. 

The problem was not that Einstein did not find tensor identities
that were equivalent to Maxwell's equations, he actually found
too many of them - and nobody knows which identity is the right
one that represents Maxwells' equations - if there is any.
For several reasons (see, e.g. \citeNP{Unz:96}, section~2.7),
 the proposed {\em field equations\/}  
(\citeNP{Einst:30}, eqn.~29 and 30)
must be wrong.\fo{In a letter to Salzer (1938, published
in \citeyearNP{Sal:74}), Einstein named as a reason for the 
failure of his teleparallel theory its representation of
the electromagnetic field in first approximation,
which does not transform as a tensor. 
We shall touch this problem in section 5.3.}

This does not imply, however, that the quantities
he considered cannot have a reasonable meaning.

\subsection{The electromagnetic field}
We shall stop here as well for a moment and investigate what
differential geometric quantities Einstein proposed for
the electromagnetic field. 
In first approximation, he defines the electromagnetic field
$a_{a \m}$ in (\citeNP{Einst:30}, eqn.~45) as
\be
a_{a \mu}= \bar h_{a \mu} -\bar h_{\mu a},
\label{emfield}
\ee
the antisymmetric part of the vielbeins $\bar h_{\mu a}$.
The vielbeins $h$, as we shall see below, are nothing other than 
a generalization of the deformation gradient in continuum mechanics. 
In the case of a compatible deformation, the antisymmetric part defined
in (\ref{emfield}) is just the curl of the displacement vector ${\bf u}$ -
the same quantity\footnote{Actually, it is not clear from Einstein's paper
whether he considered the $a_{a \m}$ as the tensor of the electromagnetic
field or its dual ($\vec E$ and $\vec B$ interchanged) - for Maxwell's equations
in empty space it makes no difference.} that had been proposed by MacCullagh~!

Thus, this part of Einstein's proposal was a kind of recycling MacCullagh's old
idea - I don't know if he was aware of that
and if he had liked it, if he were.
It seems that at that time Einstein had given up denying 
the existence of an aether \cite{Kos:92}, but probably not because 
he was aware of that relation to MacCullagh's theory.

Einsteins theory, however, is in a sense more general than MacCullagh's -
Einstein's continuum cannot be described by a compatible deformation
generated by a displacement field; this is a consequence of the nonvanishing
torsion.

We will see that there remains a close relation between Einstein and MacCullagh
as well. For this, a little excursion is needed to understand what torsion means.

\fussy
\subsection{Dislocations - a tool to understand torsion}
\sloppy
\begin{figure}[h]
\includegraphics[width=80mm]{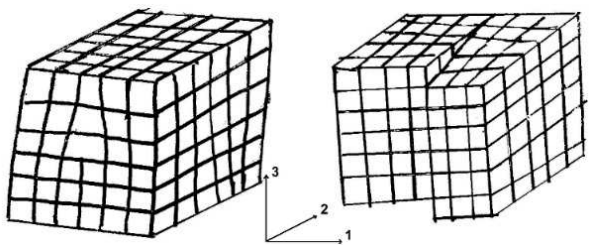}
\caption{Examples of an edge (left) and a 
screw (right) dislocations in a crystal.}
\label{disloc}
\end{figure}

\label{tool}

In \citeyearNP{Kon:52} Kondo revealed in an article of his wonderful review
`RAAG memoirs - the unifying study of basic problems in physics and
engeneering by means of geometry' the relation between dislocations
 and torsion. He discovered that torsion could be identified with a density
of dislocations piercing through a surface element.
The various components of torsion can be visualized in the example
Fig.~\ref{disloc}, an edge dislocation (left) and a 
screw  dislocation (right) in a crystal. Suppose direction 1, 2, 3
point to the right, backwards and up as indicated. Then
in the left picture, after surrounding a surface element
in the 1-3-plane one gets shifted in direction 1, therefore
this gives a contribution to the $T_{13}^1$ component of the 
torsion tensor. In the right picture, after surrounding
a surface element in the 1-2 plane the shift is in direction
3, therefore this contributes to the $T_{12}^3$ component.
Note that the singularity line of the dislocation in the 
left case goes in direction 2 and is perpendicular to 
the shift (Burgers vector),
and in the right case parallel to the shift (both in 
direction 3). Torsion is just a continuous version of 
dislocation density, that means one lets the
 lattice spacing
go to zero while maintaining the quantity shift per surface 
element. 

\citeN{Bil:55} have observed this equivalence independently
and Kr\"oner (\citeyearNP{Kro:59}; \citeyearNP{Kro:60})
made a beautiful theory out of 
it.\footnote{Kr\"oner was fascinated by the 
similarities of this geometry and wrote: " We  have  seen that
Riemannian   geometry  was  too  narrow  to  describe  dislocations  in
crystals.  Is  there  a reason why space--time has to be described by a
connection  that  is  less general than the general metric--compatible
affine connection~?" (\citeNP{Kro:60}, par.~18)}

Now, what can we learn from that~?
On the one hand, that differential geometry with torsion is
a good tool for describing dislocation behaviour in crystals.

On the other hand, we are able to give a physical interpretation
to abstract geometries like those proposed by Einstein.
In particular, there is no need to stick to the notion of a 
continuous torsion field. Spacetime could as well be endowed
with a discrete torsion on a microscopic level that appears
as dislocation density on the large scale. In this case it could be
described by a compatible displacement field ${\bf u}$ which 
has, however, singularities. Dislocations can be seen as 
singularities with Dirac-delta-valued torsion, but not all
singularities in an elastic solid need to be dislocations. 
In section \ref{Cha} I will discuss a topological defect that 
carries torsion without being a dislocation.

\subsection{Cartan and topology}
To be fair, one must say that Einstein did exclude that 
possibility and postulated a priori singularity-free
solutions. Cartan, \fo{The interesting discussion between
Einstein and Cartan is cited in the book by 
\citeN{Deb:29}.} however, told him that
postulating singularity-free solutions may create
topological complications:
\bq
`As far as {\em singularity-free\/} solutions are
concerned, it seems to me, the question is extremely 
difficult. (...) It is quite possible that the existence
of singularity-free solutions imposes purely topological
conditions on the continuum. (...)
The space in which the group exists, therefore
depends {\em from the topological point of view\/},
on the constants $\L_{ij}^{k}$ (the torsion tensor),
 and every choice of the 
constants gives a space (or family of spaces)
which is topologically defined. In short,
{\em every singularity-free solution of system 
(1),\/}\fo{The equation $\L_{\a \b ; \m}^{\g} =0$, whereby
$\L_{\a \b}^{\g}$ is the torsion tensor.}
{\em creates from the topological point of view\/}
the continuum in which it
exists'. (letter to Einstein dated Jan 3rd, 1930)
\eq
Unfortunately, Einstein was not very interested\footnote{For
a discussion of these topics, in particular the
Einstein-Cartan correspondence  see also the papers by
Vargas (\citeyearNP{Var:91a}; \citeyearNP{Var:97a}; \citeyearNP{Var:99}).}
in the topological issues that
arise in geometries with nonvanishing torsion:
\bq
`I cannot tell anything about the connectivity properties of space,
but it seems unavoidable to demand singularity-free solutions...'
(letter to Cartan dated Jan 30th, 1930)
\eq

Einstein's theory, however (or, in general, theories with torsion),
allows an interpretation as geometry with a density
of singularities on the microscopic level.

What do we gain with speculating about a discrete version
of torsion and the interpretation as topological defects~?
I consider this interesting because it establishes a connection
to a theory of physics that has been considered to be in 
blatant contradiction to Einstein's unified field theory - 
quantum mechanics. We shall see this in the following.

\section{Topological defects} \label{QB}
\subsection{Quantum behaviour}
Consider  a pair of edge dislocations (as shown
in fig.~\ref{disloc}~a) 
in a two-dimensional view fig.~\ref{2dis}. 
\begin{figure}[h]
\includegraphics[width=80mm]{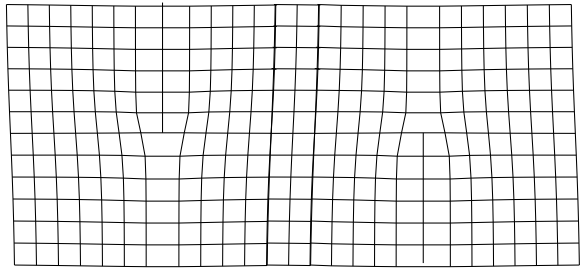}
\caption{Two edge dislocations of opposite sign in a
 two-dimensional view. Their existence cannot be deduced by
counting lattice points on the boundary as in the case of 
one single dislocation. If the two dislocations move towards 
each other, they will annihilate in the center.}
\label{2dis}
\end{figure}
It is clear that to every such defect exists an antidefect (in
this case, with the Burger's vector pointing in the opposite
direction). If the two dislocations of opposite sign in the 
left and the right part of the picture start propagating
towards each other, there will be an annihilation in the center.
No topological irregularity of the lattice will be maesurable,
even if the elastic energy stored before will give rise to 
some lattice waves. If two dislocations as those in Fig.~\ref{2dis}
move towards each other with a given velocity, it is even
conceivable 
 that the  annihilation energy creates two other defects
- not necessarily of the same structure. Thus, sticking to
the particle picture, the encounter could be even seen as a
scattering process, or it could appear as if the two dislocations
pass through each other without interacting.

This doesn't seem extraordinary at all, but has some
 noteworthy consequences if we compare the motion 
of these defects with 
the motion of classical particles. 

Fig.~\ref{feynm} a) shows the motion of a single dislocation 
propagating in $x$-direction
from point $P$ to $Q$. The slope in the $x-t$ diagram is a measure
of its velocity.
\begin{figure}[hbt]
\includegraphics[width=75mm]{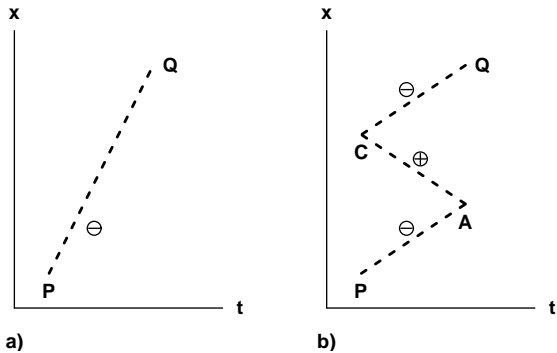}
\caption{The propagation of dislocations is analogous
to propagating electrons. When measuring the events in $P$ and $Q$,
there is no method to
detect whether a defect propagating from  $P$ to $Q$ goes a `direct'
path (a) or has a creation--annihilation process plugged in between (b). 
The `Feynman diagrams' (a) and (b) are indistinguishable.}
\label{feynm}
\end{figure}
Analogously, fig.~\ref{feynm} can be interpreted
as a Feynman Diagram for an electron propagating 
from $P$ to $Q$ (\citeNP{FeyQED},
p.~99 and p.~125).
The two signs of the dislocations correspond to
 the two signs of an electron and a positron; the latter one may be
seen as an electron travelling backwards in time.
 
If one measures only the events $P$ and 
 $Q$, besides the `direct path' Fig.~\ref{feynm}~(a) 
the scenario (b) is possible as well: while propagating, 
the dislocation encounters its 
antidefect created by a spontaneous pair creation process and 
cancels out, whereas the other `half' of the pair, identical
to the original defect, continues propagating. 

Of course, there may
be many other scenarios with the same experimental outcome
corresponding to the various Feynman graphs 
(see \citeNP{FeyQED}, p.~125).

Obviously,
it doesn't make sense to assign an `identity' 
 to this kind of `particles'. Once two defects have the same structure, they 
are identical. This kind of behaviour is well-known
in quantum mechanics. Since particles are indistinguishable, 
they have to be described by Bose-Einstein or Fermi-Dirac
statistics rather than by the classical Maxwell distribution.

Furthermore, it is clear that it makes no sense to speak about
a `trajectory' of the dislocation. This reminds us from the
result of the double slit experiment that tells us that
 it makes no sense to say the 
electron passed through the one slit or the other. 

Hence, dislocations behave not only relativistically, but also as 
quantum mechanical particles. Einstein, who introduced a geometry
that describes dislocations may have been closer to the
discovery of the puzzling quantum behaviour as he liked. 

\subsection{Homotopic classification of defects}

In the above sections we have seen
 some examples of topological defects.
For a precise definition of topological defects and for their
classification, homotopy theory is needed.
\begin{figure}[h]
\includegraphics[width=60mm]{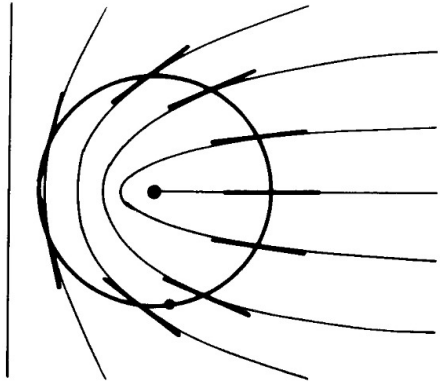}
\caption{Topological defect in a nematic liquid crystal corresponding
to the nontrivial element of the first homotopy 
group of the order parameter group $RP^2$. The circle represents
a path in $RP^2$ that cannot be contracted.}
\label{nem}
\end{figure}
 It has been applied first
by \citeN{Tou:76} to defects in ordered
 media, good review articles are
\citeN{Rog:76}, \citeN{Min:80},
 \citeN{Mic:80}, \citeN{Dzy:80},
\citeN{Mon:93} and \citeN{Nak:95}, chap.~4.8.
Ordered media have a so-called order parameter, in the example of 
fig.~\ref{nem} the orientation of the bars
in a nematic liquid crystal\fo{A fluid consisting of little rods.}.
Since this orientation is described by a 
{\em director\/}, the corresponding
{\em order parameter space\/} is the 
projective plane $RP^2$, which describes
all possible orientations of these directors in space.
In three dimensions, line defects (like the example fig.~\ref{nem})
can be detected by surrounding them
by closed lines (homeomorphic to the circle 
$S^1$) and point defects by surfaces
(homeomorphic to the sphere $S^2$). A 
topological defect occurs if the
line or surface, mapped to the order
 parameter space, cannot be contracted
to identity. Consequently, the first 
homotopy group consists of 
noncontractible loops in the order 
parameter space, and the second 
homotopy group to noncontractible closed 
surfaces. 
Nontrivial examples of the third homotopy
 group are more difficult to visualize,
a quite comprehensive example (the 
Shankar monopole) is given by \citeN{Nak:95}.
The dislocations discussed above are 
line defects. The order parameter is
the position of the 3-bein. Since 
after surrounding a defect this 3-bein
is never rotated, the homotopy group 
consists of the possible shifts
(by the discrete amount of the Burgers 
vector) in 3d-space, which is $\Z
\times \Z \times \Z$. Later I shall investigate the homotopy 
groups of $SO(3)$, the rotations in three-dimensional space.

\section{Nonlinear continuum mechanics and spacetime analogies}
\label{NC}

\subsection{Basic concepts}
It has been known for a long time that the governing
equations (\ref{navier}) of elasticity theory in 
the incompressible case 
lead to Maxwell's equations in empty space.
As far as elasticity is considered, these equations are
no more than a rather crude approximation - the classical
linear theory that applies to small deformations. 
If one wants to push further the analogies between
an elastic continuum and spacetime, there is no
physical reason to assume the deformations to be
small, in particular in the neighbourhood of
topological defects which - in the view of section~2 - 
should take the part
of elementary particles.
Therefore, there is a quite natural option to
generalize electrodynamics to a nonlinear
theory\footnote{We should not forget the
inconsistencies of classical linear electrodynamics mentioned
above. Various attempts to modify electrodynamics are described
in \shortciteN{FeyII}, chap. 28.5.} - just see how the real, nonlinear
physics of elastic continua works.

\paragraph{Deformation gradient.}
Nonlinear elasticity goes back to the work of
Cauchy in \citeyear{Cau:827}. 
One assumes an undeformed, euclidean
continuum with Cartesian coordinates ${\bf  X}= (X, Y, Z)$
(the so-called `reference configuration') and attaches
in every point a displacement vector ${\bf u}$ that points
to the coordinates ${\bf x} = (x, y, z)$ of the deformed 
state (`configuration'): ${\bf u} = {\bf x} - {\bf X}$.
\footnote{I use the common notion, e.g. in \citeN{Tru:60},
par.~15 ff.,
\citeN{Tru:65}, or, in a more modern and didactical
introduction, \citeN{Bea:87}.}

From ${\bf u}$ one deduces the basic quantity in continuum 
mechanics that transforms the coordinates ${\bf X}$
of the undeformed state to those ${\bf x}$ of the
deformed state: the so-called deformation gradient
\bd
{\bf F} := \frac{\partial {\bf x}}{\partial {\bf X }},
\ed
or, in components,
\be
{\bf F}: = \left( \begin{array}{ccc}
1+u_X &  u_Y & u_Z \\
v_X & 1+v_Y & v_Z \\
w_X & v_Y &  1+w_Z\\
\end{array} \right),
\label{F}
\ee
where $(u,v,w)$ denote the components of ${\bf u}$ and
the subscripts differentiation.

It was Cauchy's merit to discover the importance
of the symmetrical tensor
\be
{\bf B} : = {\bf F F^T},
\label{cauchy}
\ee
which is now called {\em right\/} Cauchy-Green 
tensor.\fo{${\bf C:= F^T F}$ is called the 
left Cauchy-Green tensor.}

\paragraph{Polar decomposition.}
In tensor calculus it is a very common operation
to split tensors in a symmetric and skew-symmetric part.
In continuum mechanics, however, it would not make much sense 
to split the deformation gradient tensor in that way, because
one could not assign the physical meaning of a deformation
to the results any more. The above splitting would be a linear
operation based on addition of matrices.

However, if a material undergoes a deformation ${\bf F_1}$ 
and then a successive deformation ${\bf F_2}$, 
one has to {\em multiply\/} the matrices ${\bf F_1}$ and ${\bf F_2}$
to obtain the result ${\bf F_{12}}$ - which is 
a noncommuative operation. Only for the small deformations
(a case which is frequently assumed in linear elasticity)
 one can add (${\bf F_1 - E}$) and 
(${\bf F_2 - E}$) (${\bf E}$ is the unit matrix) and get an
approximate result ${\bf F_{12} - E}$.

Fortunately, mathematicians have developed the right way to
split ${\bf F}$ multi\-plica\-tative\-ly- it's 
the {\em polar decomposition\/}\fo{${\bf F}$ has to be nonsingular.}
of 
\bd
{\bf F} = {\bf R U} =   {\bf  V R},
\ed
where ${\bf U}$ and ${\bf V}$\footnote{This decomposition is unique.
While one can find a proof of this theorem in many books, it is
rarely told how to do it in practice.  
Surprisingly, there is no way to express the
coefficients of ${\bf R}$ and ${\bf U}$ by the $f_{ij}$ in 
general. Since the solution of this problem involves the zeros
of the characteristic polynom of the matrix, the general formula
of {\sc Cardano\/} makes the solution inhibitively complicated
even for computer algebra systems.}
are positive, symmetric matrices 
and ${\bf R}$ is a rotation matrix, that means an element
of the special orthogonal group $SO(3)$. 
 Of course, a product
in the above equation means matrix multiplication, and ${\bf V}$
and ${\bf U}$ are in general different because of their
 noncommutativity.

\paragraph{Differential geometry.}
Even if there is very little overlap between the
languages of books on elasticity and books on
differential geometry\footnote{An exception 
is  \citeN{Mar:83}.}, one should be aware of the
similarity of some concepts. The deformation 
gradient ${\bf F}$ corresponds to the basis
1-forms $\th^i$ (cfr. \citeNP{Nak:95}, section~7.8). The
difference is just that the $\th^i$ maintain their 
meaning as quantities that transform coordinates,
even if they cannot be deduced from a displacement
field ${\bf u}$ any more. Since $g_{\m \n} = 
\th_{\m}^{i} \th_{\n}^{i}$,\footnote{see also
\citeNP{Einst:28}, eq.~(3), \citeNP{Einst:30}, eq.~(7).} 
is equivalent to eqn.~(\ref{cauchy}), the Cauchy tensor 
${\bf B}$ acquires the meaning of a metric. The so-called
{\em compatibility\/} conditions which are necessary 
for the existence of a displacement field ${\bf u}$
(see \citeNP{Var:93} or \citeNP{Unz:96}),
are expressed in differential geometry by the
vanishing of both the curvature and the torsion tensor.

Since the Cauchy tensor ${\bf B}$ and the stress
tensor ${\bf T}$ are both symmetric and coaxial,
the meaning of a metric can be assigned to the latter
tensor, too.
An interesting comment on the duality of 
these tensors was given by \citeN{Kro:86}.
Despite of the elegance of differential forms
I do not know, however, how to express the above
polar decomposition essential for nonlinear elasticity
properly in that language.

\subsection{Nonlinear extension of MacCullagh's 
proposal}

Before going ahead, let's keep track   
of the quantities MacCullagh and Einstein had
proposed for the electromagnetic field. 
The components of MacCullagh's $curl \ {\bf u}$ or
\bd
\partial_i u^k -\partial_k u^i
\ed
are just the skew-symmetric part of the deformation
gradient tensor ${\bf F}$ (see eqn.~\ref{F}).
The same holds for the 
the antisymmetric part of the vielbeins $\bar h_{\mu a}$ 
(eqn.~\ref{emfield})
Einstein proposed for the electromagnetic field. 

Einstein was aware that this could be an approximation only, 
but he did not put in question the process of splitting
the $h_{\mu a}$'s in a symmetric and skewsymmetric part.
From MacCullagh's point of view, however, it would have been 
a rather natural option to pass from linear elasticity to
the more general nonlinear equations.

Thus, if one wants to develop the analogy between spacetime
and an elastic continuum as consequent as possible, one
should apply to the vielbeins $h_{\mu a}$ or the 
deformation gradient ${\bf F}$ the polar decomposition theorem
and identify the electric field with the rotational part
${\bf R}$, which takes values in $SO(3)$.\fo{For small fields,
this is equivalent to MacCullagh's proposal $curl \ {\bf u}$.}

According to MacCullagh, the magnetic field corresponds to
the velocity of the aether elements (\citeNP{Whi:51}, pp. 142, 280).
 Taking this into account,
in the proposed nonlinear extension, the electromagnetic field
takes values in $SO_0(3,1)$, the connected component of the
Lorentz group\footnote{The velocity of the aether material,
 according to special relativity, can be seen as a 
pseudorotation in the $x-t$-plane.}.
Since this  does not create further topological
complications, I will sometimes restrict the discussion
to $SO(3)$ in the following. 

As mentioned above, MacCullagh's theory did not allow charges
because of the vector analysis rule $ div \ curl = 0$,
applied to the displacement field ${\bf u}$. The nonlinear extension
of MacCullagh's idea will help to overcome this difficulty
by introducing a topological defect that acts as a `source' 
of the rotations.

\subsection{Topological defects as charges in MacCullagh's theory}

\paragraph{Circular edge disclination.}
\label{Cha}
If we assume spacetime locally to be described by the deformation
gradient ${\bf F}$, by the polar decomposition theorem 
$ {\bf F= R \ U}$
follows that a unique field ${\bf R}$   can be assigned to every
point of spacetime.\footnote{If this is the case, a teleparallel connection
can be defined, see also \citeN{Tru:65}, par.~34.}
In the language of the homotopic description
of topological defects, one may regard ${\bf R}$, which takes values
in $SO(3)$, as an order parameter field and ask about possible
defects. Since the first homotopy group $\pi_1(SO(3))$ is $\Z_2$, the group 
with two elements, mathematics predicts the existence of a line defect.
The defect is then a closed
singularity line, because defect lines cannot end inside the medium.
In an elastic solid, this defect can be realized as follows:
\label{macC}
\begin{figure}[h]
\includegraphics[width=90mm]{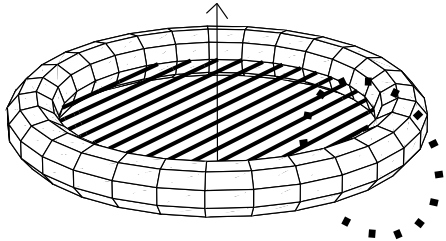}
\caption{Scematic description how to produce the topological
defect 1 $\in \pi_1(SO(3))=\Z_2$ in an elastic continuum. The solid
torus is removed. Then the material is cut along the hatched 
surface. After twisting the cut faces by the amount of 2$\pi$, the material
is rejoined. The dotted line represents a closed path in $SO(3)$ which is
not contractible. To obtain a line defect, the solid torus can be shrunk to
a singularity line. Note that after the cut, the same material elements
are rejoined.}
\label{sing}
\end{figure}

One cuts the continuum along a (circular) surface,
twists the two faces against each other by the amount of 2$\pi$
and rejoins them  again  by gluing \cite{Unz:96}.\footnote{The 
deformation gradient takes then infinite values at the boundary
of this circular surface.}

An alternative way to describe the same process is (see fig.~\ref{sing}):
Imagine $\R^3$ filled with elastic material and remove a solid torus
centered at the origin and with $z$ as symmety axis (fig.~\ref{sing}).
Then the complement\footnote{For topologists, the complement
is a solid torus as well, if $\R^3$ is compactified by adding the 
infinitely distant point. Two solid tori, merged in the described 
way, form the three-dimensional sphere $S^3$.}
is double connected due to the `bridge'
along the $z$-axis. One cuts now the material along the 
surface bounded by the inner circle of the torus in the $x-y$-plane 
(hatched surface in fig.~\ref{sing}). Now the cut faces may be twisted, 
for example clockwise the face of 
the positive $z$-direction and counterclockwise the face of the negative
one, and glued together again. If each of the twists amounts to 
$\pi$, the material elements meet their old neighbours again,
so to speak, since the total twisting angle is 2$\pi$.
To come back to the above description, now 
let the removed torus become infinitely thin.

A defect of this kind has been investigated first by \citeN{Mur:70},
who called it edge {\em disclination\/} loop. Physically, it is more
similar to a screw {\em dislocation\/}, even if closed loops
of this kind do not exist \cite{Unz:96}. Mura considered general 
twisting angles (Frank vector's), whereas for topological
reasons the twisting angle of 2$\pi$ is the most interesting.\footnote{
Any twisting angle which is not a multiple 
of 2$\pi$ would destroy the distant parallelism, i.e. the vielbeins
could not be defined any more. Accordingly, disclination density
is usually described by a nonvanishing curvature tensor.}
To come back to homotopy theory, immagine now a closed path 
in the complement (the dotted line in fig.~\ref{sing}) 
that surrounds the removed torus , e.g. the 
singularity line. A twist of 2$\pi$ in $SO(3)$ is
equivalent to identity, thus the path is closed in $SO(3)$ as well.
Since it is not contractible, the defect corresponds to the 
nontrivial element 1 of the first homotopy group $\Z_2$.
This topological description of the continuum defect has been
given first by \citeN{Rog:76}.

\paragraph{Gaussian surface integral.}

We shall see now that the defect described above can be seen
as an electrical charge in the nonlinear extension of MacCullagh's
theory. 
Except at the singularity line, a continous field of the 
deformation gradient ${\bf F}$ is given everywhere. Consequently,
also the field ${\bf R}$ obtaind by polar decomposition is
continous, since in the region of the cut-and-glue surface the 
clockwise and the counterclockwise twist by the amount of $\pi$
are the same element of $SO(3)$.

As the electrical field $\vec E$, 
one could regard an element of
$SO(3)$ as having a `direction' $(\th, \p)$ (a point
on the two-dimensional sphere $S^2$ which determines
the rotation axis), and a `length' $r$, the rotation angle\fo{this is
sometimes called an `axial vector'.}
($0 \leq r \leq \pi$ and a direction in 3D-space $0 \leq \th \leq \pi$;
$0 \leq \p \leq 2 \pi$)\footnote{This is called a representation of 
an abstract group like $SO(3)$. Another representation
uses Euler angles \cite{Lov:27}.}. 
However, even if this object has direction and length, it is not
really a vector, since `vectors' with opposite directions 
and length $\pi$ are the {\em same\/} object.\footnote{It should 
further be noted that this `vector'
has {\em no\/} `components', since due to the noncommutativity
of $SO(3)$ the superposition principle does not hold.}
This detail of replacing $curl \ {\bf u}$ by $\vec r$
will avoid the consequence $\int \!\!\!\int curl \ {\bf u} \ df=0$
and make charges possible.

\begin{figure}[h]
\includegraphics[width=90mm]{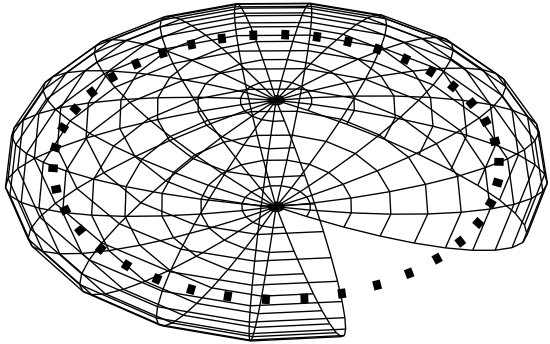}
\caption{Example of a closed surface (for a better visualization, a sector
is removed) surrounding the singularity line (dotted) generated by the
shrinking of the torus in fig.~\ref{sing}. The Gaussian integral
over this surface $f$ is considered in the following.}
\label{surf}
\end{figure}

\begin{figure}[h]
\includegraphics[width=70mm]{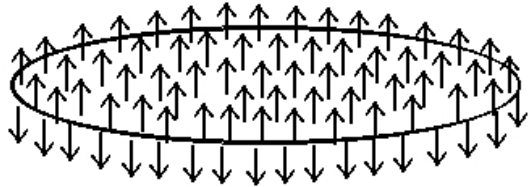}
\caption{Visualization of the Gaussian integral 
$\rho= \int \int \vec r df$ (The surface $f$ in fig.~\ref{surf}
is shrunk to a minimal size that includes the singularity).
If rotations in 3D-space a represented by a `vector'  $\vec r$, the 
rotations in the neighbourhood of the defect in fig.~\ref{sing} can
be represented as the vectors attached at the surface that contains
the singularity line. $\rho$ has nonzero value due to he fact that the
discontinous ( at z=0) `vector' field $\vec r$ represents a continous 
field of elements of $SO(3)$.}
\label{gauss}
\end{figure}

Consider  a closed surface $f$ that contains just the 
circular singularity line, like the surface shown in fig.~\ref{gauss}.
Then, in analogy to Gauss's theorem, consider the integral
\be
\r= \int \!\!\!\int \vec r \ df,
\label{gau}
\ee
where $\vec r$ denotes the `vector' in the above representation 
of $SO(3)$. If we shrink the surface $f$ in fig.~7 to a minimal
size in which it contains just the singularity line,
$\vec r$ points upwards for $z >0$ and downwards
for $z < 0$. If one integrates now {\em as if\/} $\vec r$ would be a vector,
the integral value is 2$\pi$ times the area of the circle. 

Note that
the field $\vec r$ {\em would be\/} discontinous if it were a real
vector field, but is continous, if its values are correctly interpreted
as elements of $SO(3)$.
However, the nonzero value $\rho$ of the integral appears to be a charge
if $\vec r$ is as usual identified with the electrical field $\vec E$.
This charge is a consequence of the topological properties of the
group of rotations in three-dimensional space. 
Hence, if MacCullagh's idea is consequently extended by means of
nonlinear continuum mechanics, electrically charged `particles',
i.e. topological defects become possible.

\paragraph{Quantization and screwsense.}

It is furthermore remarkable that any closed surface (like $f$ in
fig.~\ref{surf}) may contain only
integer values of these defects, in accordance with the observed
quantization of the electrical charge.

One may however raise the question how positive and negative charges
that occur in nature are distinguished in this model, since
the above defect corresponds to the element 1 of the first 
homotopy group $\Z_2$.

Abstractly, two defects of $\Z_2$ would cancel 
out by the rule $1+1=0$. It should be noted, however, that this 
defect is not described completely by the first homotopy group.
Coiled line defects like the above one do influence also
higher homotopy groups (as $\pi_2$ is influenced by $\pi_1$ of
the projective plane, see \citeN{Nak:95} and \citeN{Min:80}).
For example, the above considerations on the integral eqn.~\ref{gau}
would hold also for a defect called
 Shankar's monopole \cite{Sha:77}, representing the the
nontrivial element $1$  of the {\em third\/}
homotopy group $\pi_3(SO(3))$, which is \Z.

Apart from this it is important to note that the above defect can be 
realized in {\em two\/} different ways.
Twisting clockwise for $z>0$ and counterclockwise for $z<0$ defines
a screwsense, and if the twisting is done vice versa, the result is
not the same defect but its mirror image.
It is clear that a defect can be annihilated completely only by its
mirror image, not by an identical defect. If two identical defects
merge, the line singularity will disappear, but the result will 
be a nontrivial element of the third homotopy group $\Z$.\footnote{The 
second homotopy group is trivial.}

\subsection{Energy density in electromagnetism and continuum mechanics}

Since $SO(3)$ is compact, the electrical field could 
take finite values\footnote{Given an appropriate norm on
$SO(3)$, like the absolute value $r$ of the `vector' mentioned above.}
 only. This becomes interesting if 
one assigns an energy density to the electromagnetic
field. Then, integrating the energy density in the 
neighbourhood of an electron would not yield infinite
energies as in classical electrodynamics.

If one respects however the analogy to nonlinear elasticity theory, the 
elastic energy\footnote{We do
consider only the case in which an energy density
can be defined, which is called the case of
 {\em hyperelasticity\/}.
\citeN{Tru:65} advocate the more general case.} 
 should not depend on the rotation ${\bf R}$.
Rather it can be demonstrated that it depends on the
eigenvalues of the Cauchy tensor only - this can be deduced
by frame indifference and material 
symmetry considerations \cite{Bea:87}.

I have pointed out how MacCullagh's proposal 
of an incompressible elastic medium (\citeNP{Whi:51},
p.142) 
leads in linear approximation to
 Maxwell's equations. MacCullagh derived this by 
postulating the energy density function to 
depend on  the rotation
of the volume elements of the aether.
This additional assumption, however, is not necessary 
since Maxwell's equations already follow from the 
incompressibility condition.

The question arises if the analogy
between spacetime and an elastic continuum can be developed
as close as 
possible or if are we  forced - contrarily to the theory
of continuum mechanics - to assign 
an energy density that depends on the electric field, i.e.
to the rotation of the volume elements.

 But isn't the energy
density $w = \frac{1}{2}\e_0 (E^2 + c^2 B^2)$
of the electromagnetic field an experimental fact that
compels us to choose the latter option~?
Since this formula, together with Coulomb's law,
leads to the inconsistency of a infinite self-energy
of the electron, it is interesting to see the comment
given by Feynman on $w$:
\bq
`In practice, there are infinitely many possibilities for
$w$ and $S$ ($S$ is the Poynting vector) and up to now 
nobody has thought about an experimental method that
allows to say which  is the right one! People think that
the most simple possibility is the right one, but we must
admit that we don't know for sure which one describes the 
correct localization of the electromagnetic field energy
in space.' (\citeNP{FeyII}, chap.~27.4) 
\eq
In view of this, one should note that in a deformed elastic
solid it is obviously impossible that in a bounded region there is
a nontrivial\fo{`nontrivial' means also nonconstant here.}
 field ${\bf R}$ with a trivial Cauchy tensor
${\bf B}$\footnote{One would seek a theorem of matrix
analysis that relates the global properties of these two fields;
to my knowledge, it does not exist yet.}. Thus, the energy
density one `observes' for the electromagnetic field
could be hidden in the deformation field, as elasticity 
theory says.

\paragraph{Strain-energy function.}
Let's review briefly how nonlinear continuum mechanics
describes the localization of energy:

It is convenient to introduce the so-called
{\em principal invariants\/} $(I_1, I_2, I_3$)
of a tensor that are defined as follows:
\bea
I_1 = \l_1 + \l_2 + \l_3 \qquad(trace) \\
I_2 = \l_1 \l_2 + \l_2 \l_3 + \l_3 \l_1 \\
I_3 = \l_1 \l_2 \l_3 \qquad (det) 
\eea
(the $\l_i$ are eigenvalues).
Then, in general, the energy density $W$ is a
{\em function\/} of the principal invariants
$W = W (I_1, I_2, I_3)$ of the Cauchy tensor, and the
 stress tensor\fo{${\bf T}$ is defined as traction per surface element.
There are media with so-called
microstrucure where ${\bf T}$ is not
symmetric any more. The interested reader is referred
to the papers of \citeN{Min:64} and \citeN{Tou:64}.}
 ${\bf T}$
 is given by\footnote{This is called the
constitutive equation for isotropic materials.
 Note that there are no higher
powers of ${\bf B}$. This is a consequence of the theorem
of Cayley-Hamilton, see f.e. \citeN{Bea:87}.} 
\be
{\bf T} = \b_1 {\bf B}+ \b_0 {\bf E} + \b_{-1}  {\bf B^{-1}}
\label{stress}
\ee
with the $\b_i (I_1, I_2, I_3)$ 
being {\em functions\/} of
the principal invariants of ${\bf B}$. For small 
deformations, i.e. 
for values of the $I$'s close to 1, the
$\b$'s have fixed values - that can be
related to the known elastic 
{\em constants\/} in linear elasticity.\footnote{As
 we see, nonlinear continuum mechanics
has invented scalar-tensor-theories a
long time ago. In Brans-Dicke 
theory \citeyear{Bra:63}, the gravitational
`constant' $G$ is a function of spacetime, too.}
Of course, this complications need not to be
realized in nature, but one should keep in mind
the possibility that constants of nature 
are just the weak-field limit of
field-dependent functions.

In conclusion, the nonlinear extension of
the space\-time - elastic continuum analogy 
has interesting consequences for the localization of energy 
in space. This has to be developed and clarified
further, but may become a possibility to overcome the
inconsistencies classical electrodynamics still
has to face.

Stopping these considerations at a necessarily incomplete
stage, I'd like to mention the theoretical peculiarities
that another element of MacCullagh's theory has as
consequence - incompressibility.

\subsection{Nonlinear theory of incompressibility}
Notwithstanding its beauty, nonlinear elasticity
has suffered from the fact that, due to its
complexity, the calculation of concrete problems
was inhibitively difficult.
A big progress towards this direction was done
in the late 1940s by \citeN{Riv}, who 
 - starting from rather practical problems
given to him by the British rubber producer's
association he worked for - developed
many results of theoretical
importance (\citeNP{Riv}). 

The nonlinear condition of incompressibility
is given by 
\be
I_3 = \l_1 \l_2 \l_3 = det \ {\bf B} = (det \ {\bf F})^2 =1. \label{incomp}
\ee
and {\em not\/}, as many texts on linear elasticity state,
$div \ {\bf u = 0}$. 
As a consequence, the elastic energy $W$ depends on the 
principal invariants $I_1$ and $I_2$ of  ${\bf B}$
only. 

Rivlin considered 
 an energy density of the form
\be
W(I_1, I_2) = C (I_1 - 3) + D (I_2 -3)
\label{MR}
\ee
which is the definition of a {\em Mooney-Rivlin\/} 
material\footnote{From incompressibility follows $I_3 = 1.$}
with the constants $C$ and $D$. One should note that this
kind of material is in a certain sense the most simple 
among the isotropic incompressible ones,\footnote{Of course,
one may further assume $C$ = $D$, a case which is called 
Neo-Hookean.} but has still {\em two\/} elastic constants,\fo{It
 is not difficult to understand  why incompressible 
nonlinear elasticity 
needs more than one `shear modulus'. The reason is that squeezing and 
stretching are qualitatively different deformations. Consider a cube made
of an incompressible elastic material under pressure on the 
$x-y$ faces. It will shorten in $x$-direction and, in order
to preserve its volume, elongate in the $x-$ and $y$-
directions. In linear elasticity, the elastic energy depends
just on the ratio of the length change 
in $z$-direction, which is the same for elongation 
and shortening. For large deformations, f.e. squeezing to
the half of the height or doubling it by stretching, the
nonvanishing components of the deformation gradient would be
$f_{xx}= \l_1 = f_{yy} = \l_2 = \sqrt{2}, f_{zz} = \frac{1}{2}$ 
(squeezing) or 
$f_{xx}= \l_1 = f_{yy} = \l_2 = \frac{\sqrt{2}}{2}, f_{zz} = 2$ 
(stretching). The principal invariant $I_1$ (trace)
of the Cauchy tensor ${\bf B} = {\bf F} {\bf F^T}$ is therefore 
$2 +2 \frac{1}{4}=4  \frac{1}{4}$ in the first case and 
$\frac{1}{2} + \frac{1}{2} + 4 = 5$ in the latter
(vice versa for $I_2$). 
Thus, nonlinear elasticity distinguishes between stretching and
squeezing, or, in other words, elongation in one or two
dimensions.}
whereas in the corresponding linear case only {\em one\/}
constant, f.e. the shear modulus $\m$ describes the elastic
properties of an isotropic
incompressible material.\footnote{In linear elasticity,
the incompressibility condition is $\s=\frac{1}{2}$ (Poisson's
ratio), $k=\infty$ (compression modulus),
$\l=\infty$ (Lame's constant) or $Y= 3\m$ (Young's modulus).
See \citeN{FeyII}, chap.~38;
\citeN{Lov:27}, p.~103 for the definitions of the various constants. 
Only two of them (in the compressible case) are independent.}

\paragraph{Rivlin's cube.}
Rivlin (\citeyearNP{Riv}; \citeyearNP{Riv:74})
considered a cube of incompressible elastic material
loaded uniformely by three identical pairs of equal and
oppositely directed forces acting normally on its faces.
The surprising result was (see, f.e. \citeN{Bea:87}, p.~1719
for a derivation) that besides the trivial solution
$\l_1=\l_2=\l_3= 1$ there are six (!) others, from which
three are stable and three inherently unstable.
Hence, there might be the possibility of different stable solutions 
for the same topological defects, too.

\paragraph{Ericksen's problem.}

Determining the deformations arising from a given distribution
of body forces and surface tractions for a material with
arbitrary response functions $\b_i$ is possible for very
few simplified cases. Therefore the attention of the 
theoreticians concentrated on deformations
which arise from surface tractions alone. These defromations are
called {\em controllable\/} and are also described by the equivalent 
condition of the vanishing divergence\fo{In practice, 
this is still not easy to
calculate, since the derivations have to be taken with respect
to the deformed, curvilinear coordinate system.} of the stress tensor
$div \ {\bf T}=0$.

The reason why dealing with controllable deformations is still 
a `dirty' work is that
the material-dependent response functions $\b_i$ still enter
the calculations.
However, in nonlinear elasticity exists an interesting
class of solutions in which the $\b_i$ drop out of the 
final result, and these solutions are called `universal'.\fo{The
most common and simple example is the simple shear of a rectangular
 block that in nonlinear elasticity cannot be produced by shear
stresses only. Rather the shear stress is determined by the difference of
the normal stresses only, see \citeN{Tru:65}.}

Therefore, theoreticians were particulary interested in controllable
deformations which are independent of the $\b_i$ (universal
deformations), and \citeN{Eri:54} was the first to ask the
question: `which deformations are possible in every isotropic,
perfectly elastic body~?'.

For {\em compressible\/} materials the answer \cite{Eri:55}
is that only homogeneous deformations (that means with a constant
deformation gradient ${\bf F}$) are possible. Surprisingly, for
{\em incompressible\/} materials this is not the case. Ericksen gave 
examples of 4 different families of deformations and 
conjectured this classification to be complete. In the 
meantime, however, a fifth familiy has been detected,
and it is still an open question if there are 
others or not (see \citeN{Bea:87} and \citeN{Sac:00} for
a description of the families). 
This is just to give an example why the incompressible case is of
theoretical interest.

\subsection{Waves}
Of course, the highly nonlinear condition eqn.~\ref{incomp} 
again complicates a lot the nice behaviour of eqn.~(2)
that led to  linear electrodynamics. One important question is:
May in the nonlinear case still waves exist that correspond to the
electromagnetic waves we observe~?
Almost nothing is known about waves in the general case.

However, for the simpler case of eqn.~(\ref{MR} )
mentioned above some remarkable
results hold:
\bq
"`That is, {\em in a Mooney-Rivlin material subject to homogeneous strain,
all disturbances parallel to a given transverse principal axis are
propagated at a common speed and in unchanged form.\/} 
Perhaps this is a characterizing property of the Mooney-Rivlin material;
in any case, the possibility of waves of permanent form is certainly
unexpected in a theory of finite deformation."' (\citeNP{Tru:65},
par.~95, p.~351 above). 
\eq
Quite recently, \citeN{Bou:94} and \\
\citeN{Bou:00} have discovered
that these soliton-like solutions which maintain
their wave form while propagating exist also in the case
of arbitrary, finite deformations.\footnote{The authors 
investigated even the more general case of a previously
deformed material.}.

That means that not only the classical behaviour of
electromagnetic waves is recovered from nonlinear elasticity,
but there are hints for a particle-like behaviour of
propagating solutions coming from the nonlinear treatment.

\section{Conclusions}

The main purpose of this paper was to
attack two  popular 
preconceptions among today's physicists.

The first one regards the compatibility of aether
theories with the experimental facts of special relativity.
It has been given evidence that not the concept of the aether
as such is wrong,
but the idea of particles consisting of external material 
passing through the aether. Rather the aether is a 
concept that yields special relativity in a quite natural
way, provided that topological defects are seen as 
particles. 

Independently from this, topological defects appear
interesting, because they have been 
shown to behave as quantum mechanical particles 
under various aspects. 

The second prejudice regards the compatibility 
of quantum mechanics with Einstein' attempts
of a unified field theory using teleparallelism.
While there is no doubt that this theory presented
in the stage around 1930 is wrong, I hope to have 
convinced the reader that it is worth to be studied 
as well. 
On the one hand, there is a very close relation 
-probably unknown to Einstein- 
to the theories of the incompressible aether,
on the other hand Einstein's theory anticipated
the continuum theory of topological defects developed in the 
1950s. 
Therefore, there is a clear possibility that 
quantum theory may emerge from the geometries
Einstein considered, even if his verbal attacks
at that time still support today's common opinion of
the incompatibility of his unified theory and quantum
mechanics.

As a consequence, the author proposes to develop
the far reaching analogies between spacetime and
an elastic continuum in the most natural way-
leaving the linear approximation and apply the nonlinear
theory of finite deformations wherever possible.

The serious shortcut of MacCullagh's theory, the
impossibility of electrical charges, can be 
overcome by applying the nonlinear theory.
Other features of the
general theory, like the localization of
energy, appear promising or at least not
contradictory to experimental facts.

Even if some topics neccessarily have been discussed
in a qualitative manner, the paper should contribute
to help mathematics play a stronger role in theoretical
physics. 

In view of the open problems mentioned in section~1,
the development of the theories that have come up in 
past decades seems to be rather exhausted.

One should therefore ask the question if physics
can gain further insight from discussing events
on the Planck scale
or from the mathematics of differential
geometry and homotopy groups.

A  reconsideraton of some
old-fashioned physical theories with 
modern mathematics could be even more than
a matter of historical interest that
physicists and mathematicians of the 19th and the 
beginning 20th century merit.

\paragraph{Acknowledgement.}
I wish to thank Daniel Grumiller and Prof. Millard Beatty 
for commenting on the manuscript.

\end{document}